\documentclass[conference]{IEEEtran}
\IEEEoverridecommandlockouts

\usepackage{times}
\usepackage{soul}
\usepackage{url}
\usepackage[]{hyperref}
\usepackage{caption}
\usepackage{subcaption}
\usepackage{graphicx}
\usepackage{amsmath}
\usepackage{amsthm}
\usepackage{amssymb}
\usepackage{booktabs}
\usepackage{adjustbox}
\usepackage[square,sort,comma,numbers]{natbib}
\usepackage{orcidlink}

\usepackage{pgfplots}
\DeclareUnicodeCharacter{2212}{−}
\usepgfplotslibrary{groupplots,dateplot}
\usetikzlibrary{patterns,shapes.arrows}
\pgfplotsset{compat=newest}

\usepackage{algorithm}%
\usepackage{algpseudocode}%

\title{More Options for Prelabor Rupture of Membranes, A Bayesian Analysis}

\author{\IEEEauthorblockN{Ashley Klein\IEEEauthorrefmark{1}\IEEEauthorrefmark{2},
Edward Raff\IEEEauthorrefmark{1}\IEEEauthorrefmark{3}\IEEEauthorrefmark{4}~\IEEEmembership{Senior Member,~IEEE},
Elisabeth Seamon\IEEEauthorrefmark{2},
Lily Foley\IEEEauthorrefmark{2},
and Timothy Bussert\IEEEauthorrefmark{2}
}
\IEEEauthorblockA{\IEEEauthorrefmark{2}SUNY Upstate Medical University, Department of Obstetrics and Gynecology\\
Email: KleinAs@upstate.edu, seamone@upstate.edu, foleyli@upstate.edu, bussertt@upstate.edu}
\IEEEauthorblockA{\IEEEauthorrefmark{3}Booz Allen Hamilton\\
Email: raff\_edward@bah.com}
\IEEEauthorblockA{\IEEEauthorrefmark{4}University of Maryland, Baltimore County\\
Email: raff.edward@umbc.edu}
\thanks{\IEEEauthorrefmark{1} Ashley Klein and Edward Raff are co-first authors.}
}

\begin{document}

\maketitle

\begin{abstract}
An obstetric goal for a laboring mother is to achieve a vaginal delivery as it reduces the risks inherent in major abdominal surgery (i.e., a Cesarean section). Various medical interventions may be used by a physician to increase the likelihood of this occurring while minimizing maternal and fetal morbidity.
However, patients with prelabor rupture of membranes (PROM) have only two commonly used options for cervical ripening, Pitocin and misoprostol.
Little research exists on the benefits/risks for these two key drugs for PROM patients.
A major limitation with most induction-of-labor related research is the inability to account for differences in \textit{Bishop scores} that are commonly used in obstetrical practice to determine the next induction agent offered to the patient. This creates a confounding factor, which biases the results, but has not been realized in the literature. In this work, we use a Bayesian model of the relationships between the relevant factors, informed by expert physicians, to separate the confounding variable from its actual impact. In doing so, we provide strong evidence that pitocin and buccal misoprostol are equally effective and safe; thus, physicians have more choice in clinical care than previously realized. This is particularly important for developing countries where neither medication may be readily available, and prior guidelines may create an artificial barrier to needed medication. 
\end{abstract}

\section{Introduction}

\begin{figure}[!t]
    \centering
    \adjustbox{max width=\columnwidth}{
    \tikzset{every picture/.style={line width=0.75pt}} %

\begin{tikzpicture}[x=0.75pt,y=0.75pt,yscale=-1,xscale=1]
\draw   (250,220) -- (410,280) -- (250,340) -- (90,280) -- cycle ;
\draw   (20,474) .. controls (20,466.27) and (26.27,460) .. (34,460) -- (196,460) .. controls (203.73,460) and (210,466.27) .. (210,474) -- (210,516) .. controls (210,523.73) and (203.73,530) .. (196,530) -- (34,530) .. controls (26.27,530) and (20,523.73) .. (20,516) -- cycle ;
\draw   (260,474) .. controls (260,466.27) and (266.27,460) .. (274,460) -- (466,460) .. controls (473.73,460) and (480,466.27) .. (480,474) -- (480,516) .. controls (480,523.73) and (473.73,530) .. (466,530) -- (274,530) .. controls (266.27,530) and (260,523.73) .. (260,516) -- cycle ;
\draw   (20,80) -- (160,80) -- (160,189.09) -- (20,189.09) -- cycle ;
\draw   (180,80) -- (320,80) -- (320,180) -- (180,180) -- cycle ;
\draw    (160,130) -- (177,130) ;
\draw [shift={(180,130)}, rotate = 180] [fill={rgb, 255:red, 0; green, 0; blue, 0 }  ][line width=0.08]  [draw opacity=0] (10.72,-5.15) -- (0,0) -- (10.72,5.15) -- (7.12,0) -- cycle    ;
\draw    (250,180) -- (250,217) ;
\draw [shift={(250,220)}, rotate = 270] [fill={rgb, 255:red, 0; green, 0; blue, 0 }  ][line width=0.08]  [draw opacity=0] (10.72,-5.15) -- (0,0) -- (10.72,5.15) -- (7.12,0) -- cycle    ;
\draw    (410,280) -- (450,280) -- (450,457) ;
\draw [shift={(450,460)}, rotate = 270] [fill={rgb, 255:red, 0; green, 0; blue, 0 }  ][line width=0.08]  [draw opacity=0] (10.72,-5.15) -- (0,0) -- (10.72,5.15) -- (7.12,0) -- cycle    ;
\draw    (90,280) -- (50,280) -- (50,457) ;
\draw [shift={(50,460)}, rotate = 270] [fill={rgb, 255:red, 0; green, 0; blue, 0 }  ][line width=0.08]  [draw opacity=0] (10.72,-5.15) -- (0,0) -- (10.72,5.15) -- (7.12,0) -- cycle    ;
\draw  [fill={rgb, 255:red, 248; green, 231; blue, 28 }  ,fill opacity=0.2 ][dash pattern={on 0.84pt off 2.51pt}] (10,450) -- (490,450) -- (490,590) -- (10,590) -- cycle ;
\draw  [fill={rgb, 255:red, 189; green, 16; blue, 224 }  ,fill opacity=0.2 ][dash pattern={on 0.84pt off 2.51pt}] (10,10) -- (490,10) -- (490,200) -- (10,200) -- cycle ;
\draw  [fill={rgb, 255:red, 80; green, 227; blue, 194 }  ,fill opacity=0.2 ][dash pattern={on 0.84pt off 2.51pt}] (10,220) -- (490,220) -- (490,430) -- (10,430) -- cycle ;

\draw (90,135) node   [align=left] {\begin{minipage}[lt]{95.2pt}\setlength\topsep{0pt}
\begin{center}
Prelabor Rupture of Membranes (PROM) occurs, and timely delivery is needed
\end{center}

\end{minipage}};
\draw (250,130) node   [align=left] {\begin{minipage}[lt]{95.2pt}\setlength\topsep{0pt}
\begin{center}
Cervical exam performed and Bishop score calculated. 
\end{center}

\end{minipage}};
\draw (250,280) node   [align=left] {\begin{minipage}[lt]{129.2pt}\setlength\topsep{0pt}
\begin{center}
Bishop Score $>$ Threshold?
\end{center}

\end{minipage}};
\draw (370,495) node   [align=left] {\begin{minipage}[lt]{133.85pt}\setlength\topsep{0pt}
\begin{center}
Use buccal misoprostol.
\end{center}

\end{minipage}};
\draw (115,495) node   [align=left] {\begin{minipage}[lt]{129.2pt}\setlength\topsep{0pt}
\begin{center}
Use \ Pitocin.
\end{center}

\end{minipage}};
\draw (60,265) node   [align=left] {\begin{minipage}[lt]{40.8pt}\setlength\topsep{0pt}
\begin{center}
Yes
\end{center}

\end{minipage}};
\draw (440,265) node   [align=left] {\begin{minipage}[lt]{40.8pt}\setlength\topsep{0pt}
\begin{center}
No
\end{center}

\end{minipage}};
\draw (260,555) node   [align=left] {\begin{minipage}[lt]{244.8pt}\setlength\topsep{0pt}
All prior studies look at only this part, missing that the populations are not independent.
\end{minipage}};
\draw (255,45) node   [align=left] {\begin{minipage}[lt]{319.6pt}\setlength\topsep{0pt}
This score is not done with precise instruments, but by hand-through gloves, and thus varies. Each individual physician may have a different determination of the Bishop score’s value.
\end{minipage}};
\draw (250,385) node   [align=left] {\begin{minipage}[lt]{204pt}\setlength\topsep{0pt}
The threshold is not a hard-and-fast rule, and each physician may use their own judgment for a threshold in the intermediate range.
\end{minipage}};

\end{tikzpicture}
    }
    \caption{Diagram of the clinical process being evaluated. Patients have a cervical exam (top) that is used to calculate several measurements to form a clinically relevant Bishop score. This calculation is imprecise due to measurement error. This score is then used to determine if cervical ripening is needed (i.e., a closed cervix), at which point either buccal misoprostol or Pitocin is prescribed. Prior literature determined that Pitocin is more effective, but ignored that Pitocin was only chosen when the patient had a more favorable cervix (higher Bishop score). In this study, we include this confounding variable. }
    \label{fig:bishop_process}
\end{figure}
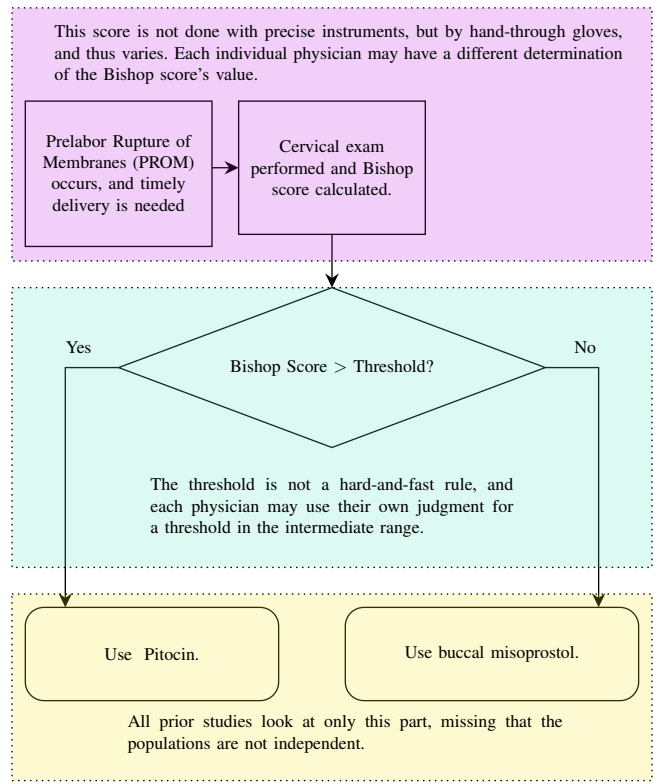

A paucity of research has occurred in women's healthcare resulting in a lack of representation of minority groups~\citep{Steinberg2023} lack of statistical rigor and power in OBGYN studies~\citep{Bruno2022}, and the lowest rate of funding by specialty with a minimal increase over a decade ~\citep{Parchem2022}. In this work, we are particularly concerned with prelabor rupture of membranes (PROM), an event that affects approximately 8\% of delivering mothers~\citep{Hannah1996}. The evidence is clear that inducing labor in PROM patients reduces complications such as intra-amniotic infection compared to expectant management. 

However, the optimal agent for induction of labor in these patients remains unclear~\citep{ACOGPracticeBulletin}. In the induction process, provider preference is often used to determine induction agents especially when there is no evidence of superiority. In practice, a cervical exam is performed with a score given to different aspects of the exam. This final tally, called a Bishop score~\citep{Bishop1964-jl}, is often used to select which medication to administer. A lower score generally indicates that the patient's cervix is ``unfavorable'' and may require a cervical ripening agent. A higher score indicates a ``favorable'' cervix which indicates that augmentation can occur with pitocin (PIT). Previous studies in PROM patients show no difference in efficacy and complications between pitocin and vaginal misoprostol. However, recent studies have shown that buccal administration of misoprostol (MISO) may be more effective and cost-effective with similar bioavailability~\citep{Alfirevic2016,Shetty2002}. 

In more detail, the test procedure is outlined in \autoref{fig:bishop_process}. A medical provider computes a Bishop score, which measures five factors by hand via a glove. If the score is above a general guideline threshold\footnote{Physicians have latitude on what specific threshold they use.}, Pitocin is recommended as the medication to use. If it is below some threshold, buccal misoprostol is used because it is a cervical ripening agent. 

There is only one prior retrospective cohort study that compares the safety and efficacy of buccal misoprostol and pitocin for PROM patients~\citep{Freret2019}. This study showed that Pitocin is more effective than buccal misoprostol with statistically significant augmentation to delivery time. However, this analysis failed to account for how \textit{the procedure of selecting the medication biases the testing of its efficacy and safety}. The common error for induction-related studies is simple statistical tests of PIT vs. MISO, without accounting for the Bishop score being a causal confounding factor. The patient was likely prescribed PIT only if they started with a more favorable cervix (higher Bishop score), and thus are already further along in their delivery. 

This study aims to disambiguate whether PIT actually results in faster deliveries and, thus, fewer adverse events than MISO. Because of other potential medications and events that may occur in delivery, it is difficult to produce a larger retrospective cohort to evaluate this empirically. Thus, we turn to Bayesian probabilistic modeling to evaluate this via a hierarchical model, where missing values are imputed as a part of the model. 

In doing so, we are able to show for the first time that Pitocin and buccal misoprostol appear to have an equal amount of time to delivery, no difference in the rate of cesarian-sections, and even equivalent numbers of patient checks. This would indicate that MISO is more viable than previously thought, and provides the evidence to consider Randomized Controlled Trials (RCTs) in the future. This is particularly important in developing countries that may not have PIT or MISO readily available at a given day or hospital~\citep{Grossman2023,Torloni2016}.

\begin{table}[!h]
\caption{A Bishop score is composed of five sub-components, each are measured by a provider. The value measured for each component is assigned an integer value, and the values are summed to produce a final score. } \label{tbl:bishops}
\centering
\begin{tabular}{@{}lcccc@{}}
\toprule
                     & \multicolumn{4}{c}{Points toward Bishop score} \\ \cmidrule(l){2-5} 
                     & 0            & 1        & 2          & 3       \\ \midrule
Dilation (cm)        & 0            & 1-2      & 3-4        & 5-6     \\
Position of cervic   & Posterior    & Mid      & Anterior   & ---     \\
Effacement           & 0-30         & 40-50    & 60-70      & $\geq$80      \\
Station              & -3           & -2       & {-1,0}     & {1,2}   \\
Cervical Consistency & Firm         & Medium   & Soft       & ---     \\ \bottomrule
\end{tabular}%
\end{table}

\section{Background}  \label{sec:background}

Due to the interdisciplinary nature of this applied work, we will first briefly review the necessary information to understand the clinical scope and applicability. This begins with the Bishop score, which has five components outlined in \autoref{tbl:bishops}. These are measured by feel, through gloves, and so precise measurements are not possible For this reason gaps in some values occur. For example, Effacement is generally measured not as a precise value (e.g., 13.2\%), but as a general range (e.g., 0-30\%). 

Pit is preferable to prescribe because it is:
1. Easier to administer. It is an infusion through the IV that is "titratable", meaning you can adjust up/down as needed. This is desirable b/c you can adjust based on baby's fetal heart tracing and number of contractions. This allows adapting to the baby and mother's tolerance of the drug to reach a more favorable outcome (vaginal delivery) over CS (surgery always has more associated risk). 

Miso is a pill, once it is given, the amount can not be altered. If baby or mom has an adverse reaction to the medication, the likelihood of a CS increases significantly. 
While other medications are available to mitigate the effect of Miso, that complicates care, increases number of potential errors, and can cause other possible adverse reactions such as postpartum hemorrhage.

To conduct our study, 7 years of patient records were reviewed by a team of three OBGYNs from 
SUNY Upsate
for all cases where PIT or MISO was prescribed, and patients were excluded if they met the following criteria: in labor, history of cesarean section, or any other condition that would preclude an attempt at a vaginal delivery. This resulted in a total of 82 patients for a retrospective study. Notably, in the data available, two of the Bishop factors, Position of the cervix and Consistency of the cervix, were combined into a single entity. Further confounding the evaluation is that this merged value was missing in 36 cases. 

With five potential outcomes of limited data with missing values, classical statistical testing is unlikely to be informative. This motivates our use of Bayesian methods to perform a hierarchical model across the related outcomes and to handle missing values in such a way that we can more reliably make conclusions about the data~\citep{Gelman2012}. 

Because patient medical data is being used, identifying information was removed by the OBGYN physicians before being made visible to the machine-learning collaborators of this study. Because this is a retrospective chart review, IRB approval is simplified to prevent any non-physician employee from observing the sensitive columns as they determine them. 

\subsection{Machine Learning Background}

Linear models are widely used in medical literature, including for meta-analysis~\cite{Klein2019}, logistic regressoions\cite{DesJardin2023}, and hierarchical Bayesian models ~\cite{DesJardin2022a}. The flexibility of Bayesian models and probabilistic programming is rarely exploited in medical studies to leverage known (or approximate) causal structures.  This work is performed in conjunction with OBGYNs to better specify the graphical model, avoiding unnecessary parameters that have no reason to inference because they are known to be unrelated to the target outcomes. Due to the sensitive nature of our dataset containing patient health information, the dataset will not be released. However, it emphasizes the importance of adapting private linear models to parity with current non-private methods~\cite{raff2023scaling,10.1145/3605764.3623910,10516654}, as the coefficients of linear models have been found sufficient to de-anonymize patients~\cite{Homer2008}.

\section{Modeling} \label{sec:modeling}

In this section, we detail the construction of our model and its design. Dealing with real-world medical health records have two primary factors in the model design choices. First is that real-world data is noisy, and in this case, we have a partially observed feature. Second, allowing us to mitigate the first, is that clinical knowledge can be used to limit and inform the scope of the model's design. 

To begin, there are three sets of dependent variables that are clinically relevant. From least to most important is \textbf{R}upture \textbf{o}f the Fetal \textbf{m}embranes (ROM), Augmentation, and Cesarian Section (CS) based outcomes. The ROM outcome has two variants for the time from a ROM event to admittance to the hospital (ROM Admit) and the time till the induction agent (PIT vs MISO) is applied. The ROM events are primarily impacted by factors outside medical control (i.e., where was the patient when ROM happened). Next is the time from augmentation with an agent (i.e., PIT/MISO started) to the time of baby delivery (Aug. Delivery) and to fully dilated (Aug. Fully) are the more important outcomes, as the intervention (PIT/MISO) has been applied. Reducing the time spent in labor (both Augmentation variables) decreases the risk of needing a CS --- the final dependent variable. CS is a major abdominal surgery, and all surgery have a higher risk of adverse events, and a longer recovery time, compared to a vaginal birth. 

For all outcome variables, we will consider the influence of each individual component of the Bishop score. However, the individual values of Positon of Cervic and Cervical Consistency are not recorded in the data. Instead, only their sum is recorded and is missing in many records. The missing value may be due to user error, lost records, urgent events preventing recording the value at the time, or other factors that are not known. 

Factoring in the Bishop components is important as it allows us to disentangle the confounding impact on the use of Pitocin vs buccal misoprostol as outlined in \autoref{fig:bishop_process}. In particular, there are cases where the individual components of the Bishop score are known to be correlated or non-important to the five dependent variables, and seeing alignment with prior known results will increase confidence in the correctness of our model.  

There are a number of input variables beyond the use of PIT/MISO and the Bishop factors. This includes:
\begin{itemize}
    \item Nullip: A ``Nulliparous'' patient is one giving birth to their first child. 
    \item Epidural: a pain reliever that is delivered via the spine neuraxially. 
    \item Fetal growth restriction (FGR): when the fetus' estimated weight or abdominal circumference is below the 10'th percentile. 
    \item Group B Streptococcus (GBS): if the patient has a specific vaginal bacteria that could be passed to the baby via vaginal delivery. 
    \item Gestational Age (GA): How far along the mother is measured in weeks 
    \item Body Mass Index: a rough ratio of a person's weight divided by their height squared. 
\end{itemize}

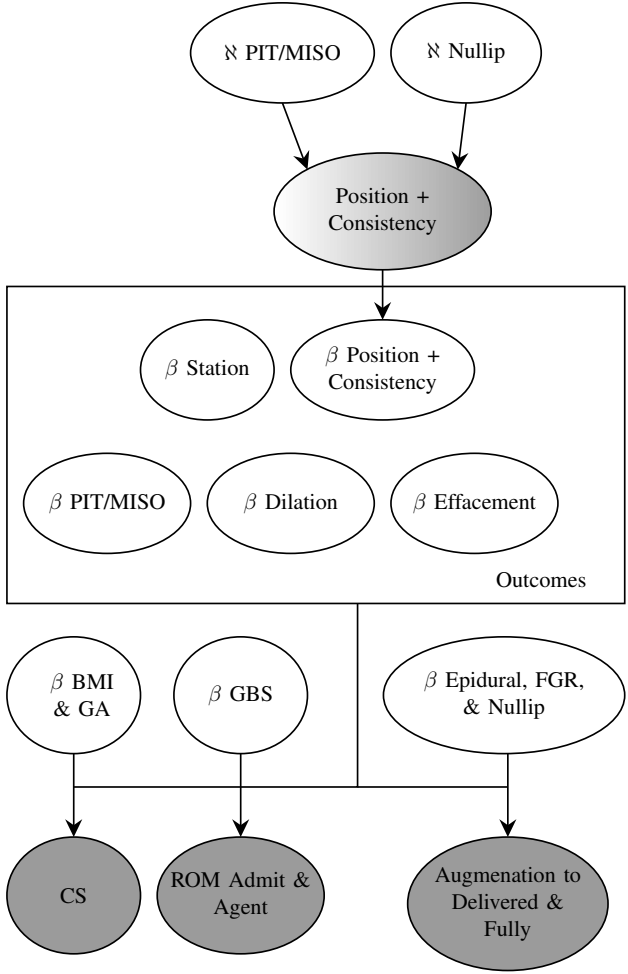
\begin{figure}[!h]
    \centering
    \adjustbox{max width=\columnwidth}{
    \tikzset {_ajgh6ahqr/.code = {\pgfsetadditionalshadetransform{ \pgftransformshift{\pgfpoint{0 bp } { 0 bp }  }  \pgftransformrotate{0 }  \pgftransformscale{2 }  }}}
\pgfdeclarehorizontalshading{_wj2cl4ede}{150bp}{rgb(0bp)=(1,1,1);
rgb(37.5bp)=(1,1,1);
rgb(62.5bp)=(0.61,0.61,0.61);
rgb(100bp)=(0.61,0.61,0.61)}
\tikzset{every picture/.style={line width=0.75pt}} %

\begin{tikzpicture}[x=0.75pt,y=0.75pt,yscale=-1,xscale=1]
\draw    (175,70) -- (188.95,107.19) ;
\draw [shift={(190,110)}, rotate = 249.44] [fill={rgb, 255:red, 0; green, 0; blue, 0 }  ][line width=0.08]  [draw opacity=0] (10.72,-5.15) -- (0,0) -- (10.72,5.15) -- (7.12,0) -- cycle    ;
\draw    (285,70) -- (280.37,107.02) ;
\draw [shift={(280,110)}, rotate = 277.13] [fill={rgb, 255:red, 0; green, 0; blue, 0 }  ][line width=0.08]  [draw opacity=0] (10.72,-5.15) -- (0,0) -- (10.72,5.15) -- (7.12,0) -- cycle    ;
\draw  [fill={rgb, 255:red, 155; green, 155; blue, 155 }  ,fill opacity=1 ] (10,545) .. controls (10,525.67) and (27.91,510) .. (50,510) .. controls (72.09,510) and (90,525.67) .. (90,545) .. controls (90,564.33) and (72.09,580) .. (50,580) .. controls (27.91,580) and (10,564.33) .. (10,545) -- cycle ;
\draw  [fill={rgb, 255:red, 155; green, 155; blue, 155 }  ,fill opacity=1 ] (100,545) .. controls (100,525.67) and (122.39,510) .. (150,510) .. controls (177.61,510) and (200,525.67) .. (200,545) .. controls (200,564.33) and (177.61,580) .. (150,580) .. controls (122.39,580) and (100,564.33) .. (100,545) -- cycle ;
\draw  [fill={rgb, 255:red, 155; green, 155; blue, 155 }  ,fill opacity=1 ] (250,550) .. controls (250,527.91) and (276.86,510) .. (310,510) .. controls (343.14,510) and (370,527.91) .. (370,550) .. controls (370,572.09) and (343.14,590) .. (310,590) .. controls (276.86,590) and (250,572.09) .. (250,550) -- cycle ;
\draw  [fill={rgb, 255:red, 255; green, 255; blue, 255 }  ,fill opacity=1 ] (180,230) .. controls (180,213.43) and (204.62,200) .. (235,200) .. controls (265.38,200) and (290,213.43) .. (290,230) .. controls (290,246.57) and (265.38,260) .. (235,260) .. controls (204.62,260) and (180,246.57) .. (180,230) -- cycle ;
\draw  [fill={rgb, 255:red, 255; green, 255; blue, 255 }  ,fill opacity=1 ] (90,230) .. controls (90,213.43) and (107.91,200) .. (130,200) .. controls (152.09,200) and (170,213.43) .. (170,230) .. controls (170,246.57) and (152.09,260) .. (130,260) .. controls (107.91,260) and (90,246.57) .. (90,230) -- cycle ;
\draw  [fill={rgb, 255:red, 255; green, 255; blue, 255 }  ,fill opacity=1 ] (240,310) .. controls (240,293.43) and (262.39,280) .. (290,280) .. controls (317.61,280) and (340,293.43) .. (340,310) .. controls (340,326.57) and (317.61,340) .. (290,340) .. controls (262.39,340) and (240,326.57) .. (240,310) -- cycle ;
\draw  [fill={rgb, 255:red, 255; green, 255; blue, 255 }  ,fill opacity=1 ] (130,310) .. controls (130,293.43) and (152.39,280) .. (180,280) .. controls (207.61,280) and (230,293.43) .. (230,310) .. controls (230,326.57) and (207.61,340) .. (180,340) .. controls (152.39,340) and (130,326.57) .. (130,310) -- cycle ;
\path  [shading=_wj2cl4ede,_ajgh6ahqr] (170,135) .. controls (170,115.67) and (199.1,100) .. (235,100) .. controls (270.9,100) and (300,115.67) .. (300,135) .. controls (300,154.33) and (270.9,170) .. (235,170) .. controls (199.1,170) and (170,154.33) .. (170,135) -- cycle ; %
 \draw   (170,135) .. controls (170,115.67) and (199.1,100) .. (235,100) .. controls (270.9,100) and (300,115.67) .. (300,135) .. controls (300,154.33) and (270.9,170) .. (235,170) .. controls (199.1,170) and (170,154.33) .. (170,135) -- cycle ; %

\draw  [fill={rgb, 255:red, 255; green, 255; blue, 255 }  ,fill opacity=1 ] (120,40) .. controls (120,23.43) and (144.62,10) .. (175,10) .. controls (205.38,10) and (230,23.43) .. (230,40) .. controls (230,56.57) and (205.38,70) .. (175,70) .. controls (144.62,70) and (120,56.57) .. (120,40) -- cycle ;
\draw  [fill={rgb, 255:red, 255; green, 255; blue, 255 }  ,fill opacity=1 ] (240,40) .. controls (240,23.43) and (260.15,10) .. (285,10) .. controls (309.85,10) and (330,23.43) .. (330,40) .. controls (330,56.57) and (309.85,70) .. (285,70) .. controls (260.15,70) and (240,56.57) .. (240,40) -- cycle ;
\draw    (235,170) -- (235,197) ;
\draw [shift={(235,200)}, rotate = 270] [fill={rgb, 255:red, 0; green, 0; blue, 0 }  ][line width=0.08]  [draw opacity=0] (10.72,-5.15) -- (0,0) -- (10.72,5.15) -- (7.12,0) -- cycle    ;
\draw  [fill={rgb, 255:red, 255; green, 255; blue, 255 }  ,fill opacity=1 ] (20,310) .. controls (20,293.43) and (42.39,280) .. (70,280) .. controls (97.61,280) and (120,293.43) .. (120,310) .. controls (120,326.57) and (97.61,340) .. (70,340) .. controls (42.39,340) and (20,326.57) .. (20,310) -- cycle ;
\draw   (10,180) -- (380,180) -- (380,370) -- (10,370) -- cycle ;
\draw  [fill={rgb, 255:red, 255; green, 255; blue, 255 }  ,fill opacity=1 ] (235.5,425) .. controls (235.5,405.67) and (267.85,390) .. (307.75,390) .. controls (347.65,390) and (380,405.67) .. (380,425) .. controls (380,444.33) and (347.65,460) .. (307.75,460) .. controls (267.85,460) and (235.5,444.33) .. (235.5,425) -- cycle ;
\draw  [fill={rgb, 255:red, 255; green, 255; blue, 255 }  ,fill opacity=1 ] (110,425) .. controls (110,405.67) and (127.91,390) .. (150,390) .. controls (172.09,390) and (190,405.67) .. (190,425) .. controls (190,444.33) and (172.09,460) .. (150,460) .. controls (127.91,460) and (110,444.33) .. (110,425) -- cycle ;
\draw  [fill={rgb, 255:red, 255; green, 255; blue, 255 }  ,fill opacity=1 ] (10,425) .. controls (10,405.67) and (27.91,390) .. (50,390) .. controls (72.09,390) and (90,405.67) .. (90,425) .. controls (90,444.33) and (72.09,460) .. (50,460) .. controls (27.91,460) and (10,444.33) .. (10,425) -- cycle ;
\draw    (50,460) -- (50,482.5) -- (50,507) ;
\draw [shift={(50,510)}, rotate = 270] [fill={rgb, 255:red, 0; green, 0; blue, 0 }  ][line width=0.08]  [draw opacity=0] (10.72,-5.15) -- (0,0) -- (10.72,5.15) -- (7.12,0) -- cycle    ;
\draw    (150,460) -- (150,507) ;
\draw [shift={(150,510)}, rotate = 270] [fill={rgb, 255:red, 0; green, 0; blue, 0 }  ][line width=0.08]  [draw opacity=0] (10.72,-5.15) -- (0,0) -- (10.72,5.15) -- (7.12,0) -- cycle    ;
\draw    (310,460) -- (310,507) ;
\draw [shift={(310,510)}, rotate = 270] [fill={rgb, 255:red, 0; green, 0; blue, 0 }  ][line width=0.08]  [draw opacity=0] (10.72,-5.15) -- (0,0) -- (10.72,5.15) -- (7.12,0) -- cycle    ;
\draw    (220,370) -- (220,480) -- (90,480) -- (50,480) ;
\draw    (220,480) -- (310,480) ;

\draw (175,40) node   [align=left] {\begin{minipage}[lt]{61.2pt}\setlength\topsep{0pt}
\begin{center}
$\displaystyle \aleph $ PIT/MISO
\end{center}

\end{minipage}};
\draw (180,310) node   [align=left] {\begin{minipage}[lt]{68pt}\setlength\topsep{0pt}
\begin{center}
$\displaystyle \beta $ Dilation
\end{center}

\end{minipage}};
\draw (235,230) node   [align=left] {\begin{minipage}[lt]{74.8pt}\setlength\topsep{0pt}
\begin{center}
$\displaystyle \beta $ Position + Consistency
\end{center}

\end{minipage}};
\draw (290,310) node   [align=left] {\begin{minipage}[lt]{68pt}\setlength\topsep{0pt}
\begin{center}
$\displaystyle \beta $ Effacement
\end{center}

\end{minipage}};
\draw (130,230) node   [align=left] {\begin{minipage}[lt]{47.6pt}\setlength\topsep{0pt}
\begin{center}
$\displaystyle \beta $ Station
\end{center}

\end{minipage}};
\draw (285,40) node   [align=left] {\begin{minipage}[lt]{47.6pt}\setlength\topsep{0pt}
\begin{center}
$\displaystyle \aleph $ Nullip
\end{center}

\end{minipage}};
\draw (150,545) node   [align=left] {\begin{minipage}[lt]{68pt}\setlength\topsep{0pt}
\begin{center}
ROM Admit \& Agent
\end{center}

\end{minipage}};
\draw (310,550) node   [align=left] {\begin{minipage}[lt]{68pt}\setlength\topsep{0pt}
\begin{center}
Augmenation to Delivered \& Fully
\end{center}

\end{minipage}};
\draw (307.75,425) node   [align=left] {\begin{minipage}[lt]{74.8pt}\setlength\topsep{0pt}
\begin{center}
$\displaystyle \beta $ Epidural, FGR, \& Nullip
\end{center}

\end{minipage}};
\draw (150,425) node   [align=left] {\begin{minipage}[lt]{46.63pt}\setlength\topsep{0pt}
\begin{center}
$\displaystyle \beta $ GBS
\end{center}

\end{minipage}};
\draw (50,545) node   [align=left] {\begin{minipage}[lt]{34pt}\setlength\topsep{0pt}
\begin{center}
CS
\end{center}

\end{minipage}};
\draw (55,425) node   [align=left] {\begin{minipage}[lt]{34pt}\setlength\topsep{0pt}
\begin{center}
$\displaystyle \beta $ BMI \& GA
\end{center}

\end{minipage}};
\draw (235,135) node   [align=left] {\begin{minipage}[lt]{74.8pt}\setlength\topsep{0pt}
\begin{center}
Position + Consistency
\end{center}

\end{minipage}};
\draw (70,310) node   [align=left] {\begin{minipage}[lt]{68pt}\setlength\topsep{0pt}
\begin{center}
$\displaystyle \beta $ PIT/MISO
\end{center}

\end{minipage}};
\draw (330,355.5) node   [align=left] {\begin{minipage}[lt]{68pt}\setlength\topsep{0pt}
\begin{center}
Outcomes
\end{center}

\end{minipage}};

\end{tikzpicture}
    }
    \caption{Plate diagram of the relationships and connections in the model. The white notes indicate latent variables and grey nodes are observed variables used for calculating loglikelihood in the model. The ``Position + Consistency'' is mixed because it is partially observed, but in other cases, its value is inferred from other variables. Covariates with $\aleph$ are used to infer the missing value, and covariates with $\beta$ are used to make the predictions. The model is hierarchical, where each $\beta$ is replicated for each outcome it is used against, with a shared hyperprior. Nodes with an ampersand (``\&'') represent multiple nodes that have the same connections but are grouped together for legibility.  }
    \label{fig:plate}
\end{figure}

Due to the limited number of samples (82), we rely heavily on practitioner guidance on which variables may plausibly be correlated with any of the four outcomes. The Plate diagram showing the overarching relationships in the model is given in \autoref{fig:plate}, where white nodes are latent variables, solid grey notes are observed outcomes, and the white-to-grey gradient is a partially observed outcome and latent variable because we must infer the value when missing in the data. 

The $\aleph$ covariates are used to predict the missing (combined) contributing Bishop score of ``Position + consistency''. Note that the $\aleph$ values are informed by both their direct likelihood calculated from the known cases, as well as the impact of the imputed score that is used in the prediction of the five target outcomes. The value is imputed from the variable PIT/MISO which is expected to be highly predictive because PIT is only given when a patient has a higher Bishop score, and thus it is more likely that Position + Consistency was higher. Whether the patient is a Nullip (first child) is also included as a known causal factor to Cervical Consistency.

Because the contributory factors to the Bishop score are integer ordinal values, we impute Position $+$ Consistency as an ordinal regression. The distribution $\pi(\cdot)$ we use is $\pi(\mathbf{c}, \mid \boldsymbol{\alpha}, \phi)
=
\mathcal{D}( \mathbf{p}(\mathbf{c}, \phi) \mid \boldsymbol{\alpha})
\cdot \left| J(\mathbf{c}, \phi) \right|
$, where $\mathbf{c}$ are the cut points for the ordinal regression, and $\mathbf{\alpha}$ and $\phi$ are parameters that control the cut location and center value, respectively. This is done with a Dirichlet distribution over $\mathbf{\alpha}$ and its Jacobian $J$ following ~\citep{betanalphaOrdinalRegression}. The likelihood is then computed using the threshold ordered logistic  ~\cite{Lu2022} \autoref{eq:orderedLogistic}. 

\begin{equation} 
\label{eq:orderedLogistic}
    \begin{split}f(k \mid \eta, \mathbf{c}) = \left\{
  \begin{array}{l}
    1 - \text{logit}^{-1}(\eta - c_1)
      \,, \text{if } k = 0 \\
    \text{logit}^{-1}(\eta - c_{k - 1}) -
    \text{logit}^{-1}(\eta - c_{k})
      \,,  \\
    \text{logit}^{-1}(\eta - c_{K - 1})
      \,, \text{if } k = K \\
  \end{array}
\right.\end{split}
\end{equation}

This results in a proper prior over the cutpoints $\mathbf{c}$, which is desirable because it allows for more inference of frequently seen values. Because all patients in this study are PROM, it is rare for them to have very high Bishop scores, and thus the highest possible score of Position + Consistency (4) is a rarer occurrence. 

The covariates denoted by $\beta$ are applied to the outcome predictors. The $\beta$'s that correspond to the Bishop scores, as well as the use of PIT/MISO, are applied to all five outcomes of interest. This is done because of the causal factor that the Bishop scores play in the determination of PIT/MISO, but has not previously been disambiguated in the literature. Our desire is to include them as explicit factors so that the impact of a patient having a more favorable cervix (high Bishop score) can be separated from the choice of PIT vs MISO in practice. 

The other variables, BMI, GA, GBS, Epidural, FPGR, and Nullip, are only applied to the outcomes that professional OBGYN physicians deemed possibly correlated. This limitation on the hypothesis space is desirable due to the small sample size to avoid endless minimization of study power.

\begin{algorithm}[!h]
    \caption{Generative Story for PROM}
    \label{alg:generative_story}
    \begin{algorithmic}[1]

        \State $\aleph_{nullip} \sim \mathcal{N}(0, 1)$
        \State $\aleph_{pit} \sim \mathcal{N}(0, 1)$
        \State $b_{pc} \sim Cauchy(0, 1)$
        \State $\eta_i \gets \aleph_{nullip} \cdot nullip_i + \aleph_{pit} \cdot pit_i + b_{pc}$
        \State $\alpha \gets \vec{1}$
        \State $\phi \sim \mathcal{U}(0,4)$ \Comment{Anchor sampled from the range of outcomes}
        \State $\mathbf{c} \sim \pi(\mathbf{c}, \mid \boldsymbol{\alpha}, \phi)$
        \State $\widehat{\text{PosCon}}_j \sim f(k \mid \eta_j, \mathbf{c})$ , See \autoref{eq:orderedLogistic}
        \For{Data points $j$ where PosCon is known}
            \State Measure likelihood $ f(k=\text{PosCon}_j \mid \eta_j, \mathbf{c})$
            \State Set $\widehat{\text{PosCon}}_j \gets \text{PosCon}_j$ \Comment{We want to use the known value when available}
        \EndFor
        \State Sample all of the covariates $\beta^\circ_{\Box} \sim \mathcal{N}(\mu_\Box, \sigma_\Box)$ where $\Box$ is a stand-in for (Dilation, Effacement, Station, PosCon), and $\circ$ is a stand-in for the outcome (ROM Admit, Rom Agent, Aug. to fully, Aug. to delivery). 
        \State $\mu^\circ_\mathit{bishop} \gets \beta^\circ_\mathit{PosCon} \widehat{\text{PosCon}} + \beta^\circ_\mathit{dilation} \text{Dilation} + \beta^\circ_\mathit{Effacement} \text{Effacement}  + \beta^\circ_\mathit{station} \text{Station} $
        \State $\hat{y}^\mathit{ROM\_Adm} \gets \mu^\mathit{ROM\_Adm}_\mathit{bishop} + \beta^\mathit{ROM\_Adm}_\mathit{GBS} \text{GBS} $
        \State $\hat{y}^\mathit{ROM\_Agent} \gets \mu^\mathit{ROM\_Agent}_\mathit{bishop} + \beta^\mathit{ROM\_Agent}_\mathit{GBS} \text{GBS} $
        \State $\hat{y}^\mathit{ADM\_Ful} \gets \mu^\mathit{ADM\_Ful}_\mathit{bishop} + \beta^\mathit{ADM\_Ful}_\mathit{Nullip} \text{Nullip} + \beta^\mathit{ADM\_Ful}_\mathit{Epi} \text{Epidural} $
        \State $\hat{y}^\mathit{ADM\_Del} \gets \mu^\mathit{ADM\_Del}_\mathit{bishop} + 
        \beta^\mathit{ADM\_Del}_\mathit{Nullip} \text{Nullip} + \beta^\mathit{ADM\_Del}_\mathit{Epi} \text{Epidural} + \beta^\mathit{ADM\_Del}_\mathit{FGR} \text{FGR} $
        \State $\hat{y}^\mathit{CS} \gets \mu^\mathit{CS}_\mathit{bishop} + \beta^\mathit{CS}_\mathit{BMI} \text{BMI} + \beta^\mathit{CS}_\mathit{GA} \text{GA} $
        \State Observe likelihood of $\hat{y}^\mathit{CS}$ against the Bernoulli distribution against the known $\mathit{CS}$
        \State For each remaining $\hat{y}^\circ$, observe its likelihood against $\mathcal{N}(\hat{y}^\circ, \sigma^\circ)$
      \State $\beta_{CS} \sim \mathcal{N}(\mu_{CS}, \sigma_{CS})$
    \end{algorithmic}
\end{algorithm}

The more detailed generative store for the model we consider is thus presented in Algorithm \autoref{alg:generative_story}. The symbol $\mathcal{N}(a,b)$ is used to denote the Gaussian distribution of mean $a$ and standard deviation $b$, and Cauchy the Cauchy distribution. 

To avoid clutter in the algorithm, hyperpriors are implied by the inclusion of an undefined variable. For example, $\mathcal{N}(\mu, \sigma)$ would indicate a hyper-prior with $\mu \sim \mathcal{N}(0, 1)$ and $\sigma \sim HalfNormal(1)$, where ``Half''-Normal is used to indicate a normal distribution truncated to only the positive range (and re-normalized). In all cases, the location hyperprior is Gaussian and the spread hyperprior is HalfNormal, unless specified otherwise. If two variables use the exact same symbol then they share the same hyperprior. 

We also use the notation of $\circ$ and $\Box$ as wild-card patterns to indicate multiple variables initialized in the same way. This is done for Dilation, Effacement, Station, and PosCon (Position + Consistency) because each has a hyperprior that is shared over all outcome variables. The $\Box$ is used to indicate which input variables the covariate is associated with, and $\circ$ to indicate which of the outcomes have a distinct $\beta$ value via the hyperprior. 

\begin{figure}[!h]
\begin{tikzpicture}

\definecolor{darkgray178}{RGB}{178,178,178}
\definecolor{silver188}{RGB}{188,188,188}
\definecolor{steelblue52138189}{RGB}{52,138,189}
\definecolor{whitesmoke238}{RGB}{238,238,238}

\begin{axis}[
axis background/.style={fill=whitesmoke238},
axis line style={silver188},
tick pos=left,
x grid style={darkgray178},
xmajorgrids,
xmin=-0.98333333315, xmax=36.41666667015,
xtick style={color=black},
y grid style={darkgray178},
ylabel={Count},
ymajorgrids,
ymin=0, ymax=25.2,
ytick style={color=black},
width=0.95\columnwidth,
height=0.6\columnwidth,
title={Augmentation to Delivery},
]
\draw[draw=whitesmoke238,fill=steelblue52138189,fill opacity=0.75,very thin] (axis cs:0.716666667,0) rectangle (axis cs:3.80757575818182,8);
\draw[draw=whitesmoke238,fill=steelblue52138189,fill opacity=0.75,very thin] (axis cs:3.80757575818182,0) rectangle (axis cs:6.89848484936364,14);
\draw[draw=whitesmoke238,fill=steelblue52138189,fill opacity=0.75,very thin] (axis cs:6.89848484936364,0) rectangle (axis cs:9.98939394054546,24);
\draw[draw=whitesmoke238,fill=steelblue52138189,fill opacity=0.75,very thin] (axis cs:9.98939394054546,0) rectangle (axis cs:13.0803030317273,15);
\draw[draw=whitesmoke238,fill=steelblue52138189,fill opacity=0.75,very thin] (axis cs:13.0803030317273,0) rectangle (axis cs:16.1712121229091,11);
\draw[draw=whitesmoke238,fill=steelblue52138189,fill opacity=0.75,very thin] (axis cs:16.1712121229091,0) rectangle (axis cs:19.2621212140909,3);
\draw[draw=whitesmoke238,fill=steelblue52138189,fill opacity=0.75,very thin] (axis cs:19.2621212140909,0) rectangle (axis cs:22.3530303052727,1);
\draw[draw=whitesmoke238,fill=steelblue52138189,fill opacity=0.75,very thin] (axis cs:22.3530303052727,0) rectangle (axis cs:25.4439393964545,2);
\draw[draw=whitesmoke238,fill=steelblue52138189,fill opacity=0.75,very thin] (axis cs:25.4439393964545,0) rectangle (axis cs:28.5348484876364,1);
\draw[draw=whitesmoke238,fill=steelblue52138189,fill opacity=0.75,very thin] (axis cs:28.5348484876364,0) rectangle (axis cs:31.6257575788182,2);
\draw[draw=whitesmoke238,fill=steelblue52138189,fill opacity=0.75,very thin] (axis cs:31.6257575788182,0) rectangle (axis cs:34.71666667,1);
\end{axis}

\end{tikzpicture}
\hfil
\begin{tikzpicture}

\definecolor{darkgray178}{RGB}{178,178,178}
\definecolor{silver188}{RGB}{188,188,188}
\definecolor{steelblue52138189}{RGB}{52,138,189}
\definecolor{whitesmoke238}{RGB}{238,238,238}

\begin{axis}[
axis background/.style={fill=whitesmoke238},
axis line style={silver188},
tick pos=left,
x grid style={darkgray178},
xmajorgrids,
xmin=-1.0991666665, xmax=36.2824999965,
xtick style={color=black},
y grid style={darkgray178},
ylabel={Count},
ymajorgrids,
ymin=0, ymax=19.95,
ytick style={color=black},
width=0.95\columnwidth,
height=0.6\columnwidth,
title={Augmentation to Fully Dilated},
]
\draw[draw=whitesmoke238,fill=steelblue52138189,fill opacity=0.75,very thin] (axis cs:0.6,0) rectangle (axis cs:3.68939393909091,10);
\draw[draw=whitesmoke238,fill=steelblue52138189,fill opacity=0.75,very thin] (axis cs:3.68939393909091,0) rectangle (axis cs:6.77878787818182,15);
\draw[draw=whitesmoke238,fill=steelblue52138189,fill opacity=0.75,very thin] (axis cs:6.77878787818182,0) rectangle (axis cs:9.86818181727273,19);
\draw[draw=whitesmoke238,fill=steelblue52138189,fill opacity=0.75,very thin] (axis cs:9.86818181727273,0) rectangle (axis cs:12.9575757563636,11);
\draw[draw=whitesmoke238,fill=steelblue52138189,fill opacity=0.75,very thin] (axis cs:12.9575757563636,0) rectangle (axis cs:16.0469696954545,6);
\draw[draw=whitesmoke238,fill=steelblue52138189,fill opacity=0.75,very thin] (axis cs:16.0469696954545,0) rectangle (axis cs:19.1363636345455,2);
\draw[draw=whitesmoke238,fill=steelblue52138189,fill opacity=0.75,very thin] (axis cs:19.1363636345455,0) rectangle (axis cs:22.2257575736364,1);
\draw[draw=whitesmoke238,fill=steelblue52138189,fill opacity=0.75,very thin] (axis cs:22.2257575736364,0) rectangle (axis cs:25.3151515127273,1);
\draw[draw=whitesmoke238,fill=steelblue52138189,fill opacity=0.75,very thin] (axis cs:25.3151515127273,0) rectangle (axis cs:28.4045454518182,1);
\draw[draw=whitesmoke238,fill=steelblue52138189,fill opacity=0.75,very thin] (axis cs:28.4045454518182,0) rectangle (axis cs:31.4939393909091,1);
\draw[draw=whitesmoke238,fill=steelblue52138189,fill opacity=0.75,very thin] (axis cs:31.4939393909091,0) rectangle (axis cs:34.58333333,1);
\end{axis}

\end{tikzpicture}
\begin{tikzpicture}

\definecolor{darkgray178}{RGB}{178,178,178}
\definecolor{silver188}{RGB}{188,188,188}
\definecolor{steelblue52138189}{RGB}{52,138,189}
\definecolor{whitesmoke238}{RGB}{238,238,238}

\begin{axis}[
axis background/.style={fill=whitesmoke238},
axis line style={silver188},
tick pos=left,
x grid style={darkgray178},
xmajorgrids,
xmin=-4.62583333465, xmax=114.00916670165,
xtick style={color=black},
y grid style={darkgray178},
ylabel={Count},
ymajorgrids,
ymin=0, ymax=32.55,
ytick style={color=black},
width=0.95\columnwidth,
height=0.6\columnwidth,
title={ROM to Admittance},
]
\draw[draw=whitesmoke238,fill=steelblue52138189,fill opacity=0.75,very thin] (axis cs:0.766666667,0) rectangle (axis cs:2.76388888983333,31);
\draw[draw=whitesmoke238,fill=steelblue52138189,fill opacity=0.75,very thin] (axis cs:2.76388888983333,0) rectangle (axis cs:4.76111111266667,19);
\draw[draw=whitesmoke238,fill=steelblue52138189,fill opacity=0.75,very thin] (axis cs:4.76111111266667,0) rectangle (axis cs:6.7583333355,11);
\draw[draw=whitesmoke238,fill=steelblue52138189,fill opacity=0.75,very thin] (axis cs:6.7583333355,0) rectangle (axis cs:8.75555555833333,9);
\draw[draw=whitesmoke238,fill=steelblue52138189,fill opacity=0.75,very thin] (axis cs:8.75555555833333,0) rectangle (axis cs:10.7527777811667,4);
\draw[draw=whitesmoke238,fill=steelblue52138189,fill opacity=0.75,very thin] (axis cs:10.7527777811667,0) rectangle (axis cs:12.750000004,1);
\draw[draw=whitesmoke238,fill=steelblue52138189,fill opacity=0.75,very thin] (axis cs:12.750000004,0) rectangle (axis cs:14.7472222268333,2);
\draw[draw=whitesmoke238,fill=steelblue52138189,fill opacity=0.75,very thin] (axis cs:14.7472222268333,0) rectangle (axis cs:16.7444444496667,1);
\draw[draw=whitesmoke238,fill=steelblue52138189,fill opacity=0.75,very thin] (axis cs:16.7444444496667,0) rectangle (axis cs:18.7416666725,2);
\draw[draw=whitesmoke238,fill=steelblue52138189,fill opacity=0.75,very thin] (axis cs:18.7416666725,0) rectangle (axis cs:20.7388888953333,0);
\draw[draw=whitesmoke238,fill=steelblue52138189,fill opacity=0.75,very thin] (axis cs:20.7388888953333,0) rectangle (axis cs:22.7361111181667,0);
\draw[draw=whitesmoke238,fill=steelblue52138189,fill opacity=0.75,very thin] (axis cs:22.7361111181667,0) rectangle (axis cs:24.733333341,0);
\draw[draw=whitesmoke238,fill=steelblue52138189,fill opacity=0.75,very thin] (axis cs:24.733333341,0) rectangle (axis cs:26.7305555638333,0);
\draw[draw=whitesmoke238,fill=steelblue52138189,fill opacity=0.75,very thin] (axis cs:26.7305555638333,0) rectangle (axis cs:28.7277777866667,0);
\draw[draw=whitesmoke238,fill=steelblue52138189,fill opacity=0.75,very thin] (axis cs:28.7277777866667,0) rectangle (axis cs:30.7250000095,0);
\draw[draw=whitesmoke238,fill=steelblue52138189,fill opacity=0.75,very thin] (axis cs:30.7250000095,0) rectangle (axis cs:32.7222222323333,0);
\draw[draw=whitesmoke238,fill=steelblue52138189,fill opacity=0.75,very thin] (axis cs:32.7222222323333,0) rectangle (axis cs:34.7194444551667,0);
\draw[draw=whitesmoke238,fill=steelblue52138189,fill opacity=0.75,very thin] (axis cs:34.7194444551667,0) rectangle (axis cs:36.716666678,0);
\draw[draw=whitesmoke238,fill=steelblue52138189,fill opacity=0.75,very thin] (axis cs:36.716666678,0) rectangle (axis cs:38.7138889008333,0);
\draw[draw=whitesmoke238,fill=steelblue52138189,fill opacity=0.75,very thin] (axis cs:38.7138889008333,0) rectangle (axis cs:40.7111111236667,0);
\draw[draw=whitesmoke238,fill=steelblue52138189,fill opacity=0.75,very thin] (axis cs:40.7111111236667,0) rectangle (axis cs:42.7083333465,0);
\draw[draw=whitesmoke238,fill=steelblue52138189,fill opacity=0.75,very thin] (axis cs:42.7083333465,0) rectangle (axis cs:44.7055555693333,0);
\draw[draw=whitesmoke238,fill=steelblue52138189,fill opacity=0.75,very thin] (axis cs:44.7055555693333,0) rectangle (axis cs:46.7027777921667,0);
\draw[draw=whitesmoke238,fill=steelblue52138189,fill opacity=0.75,very thin] (axis cs:46.7027777921667,0) rectangle (axis cs:48.700000015,0);
\draw[draw=whitesmoke238,fill=steelblue52138189,fill opacity=0.75,very thin] (axis cs:48.700000015,0) rectangle (axis cs:50.6972222378333,0);
\draw[draw=whitesmoke238,fill=steelblue52138189,fill opacity=0.75,very thin] (axis cs:50.6972222378333,0) rectangle (axis cs:52.6944444606667,0);
\draw[draw=whitesmoke238,fill=steelblue52138189,fill opacity=0.75,very thin] (axis cs:52.6944444606667,0) rectangle (axis cs:54.6916666835,0);
\draw[draw=whitesmoke238,fill=steelblue52138189,fill opacity=0.75,very thin] (axis cs:54.6916666835,0) rectangle (axis cs:56.6888889063333,0);
\draw[draw=whitesmoke238,fill=steelblue52138189,fill opacity=0.75,very thin] (axis cs:56.6888889063333,0) rectangle (axis cs:58.6861111291667,0);
\draw[draw=whitesmoke238,fill=steelblue52138189,fill opacity=0.75,very thin] (axis cs:58.6861111291667,0) rectangle (axis cs:60.683333352,0);
\draw[draw=whitesmoke238,fill=steelblue52138189,fill opacity=0.75,very thin] (axis cs:60.683333352,0) rectangle (axis cs:62.6805555748333,0);
\draw[draw=whitesmoke238,fill=steelblue52138189,fill opacity=0.75,very thin] (axis cs:62.6805555748333,0) rectangle (axis cs:64.6777777976667,0);
\draw[draw=whitesmoke238,fill=steelblue52138189,fill opacity=0.75,very thin] (axis cs:64.6777777976667,0) rectangle (axis cs:66.6750000205,0);
\draw[draw=whitesmoke238,fill=steelblue52138189,fill opacity=0.75,very thin] (axis cs:66.6750000205,0) rectangle (axis cs:68.6722222433333,0);
\draw[draw=whitesmoke238,fill=steelblue52138189,fill opacity=0.75,very thin] (axis cs:68.6722222433333,0) rectangle (axis cs:70.6694444661667,0);
\draw[draw=whitesmoke238,fill=steelblue52138189,fill opacity=0.75,very thin] (axis cs:70.6694444661667,0) rectangle (axis cs:72.666666689,0);
\draw[draw=whitesmoke238,fill=steelblue52138189,fill opacity=0.75,very thin] (axis cs:72.666666689,0) rectangle (axis cs:74.6638889118333,0);
\draw[draw=whitesmoke238,fill=steelblue52138189,fill opacity=0.75,very thin] (axis cs:74.6638889118333,0) rectangle (axis cs:76.6611111346667,0);
\draw[draw=whitesmoke238,fill=steelblue52138189,fill opacity=0.75,very thin] (axis cs:76.6611111346667,0) rectangle (axis cs:78.6583333575,0);
\draw[draw=whitesmoke238,fill=steelblue52138189,fill opacity=0.75,very thin] (axis cs:78.6583333575,0) rectangle (axis cs:80.6555555803333,0);
\draw[draw=whitesmoke238,fill=steelblue52138189,fill opacity=0.75,very thin] (axis cs:80.6555555803333,0) rectangle (axis cs:82.6527778031667,0);
\draw[draw=whitesmoke238,fill=steelblue52138189,fill opacity=0.75,very thin] (axis cs:82.6527778031667,0) rectangle (axis cs:84.650000026,0);
\draw[draw=whitesmoke238,fill=steelblue52138189,fill opacity=0.75,very thin] (axis cs:84.650000026,0) rectangle (axis cs:86.6472222488333,0);
\draw[draw=whitesmoke238,fill=steelblue52138189,fill opacity=0.75,very thin] (axis cs:86.6472222488333,0) rectangle (axis cs:88.6444444716667,0);
\draw[draw=whitesmoke238,fill=steelblue52138189,fill opacity=0.75,very thin] (axis cs:88.6444444716667,0) rectangle (axis cs:90.6416666945,0);
\draw[draw=whitesmoke238,fill=steelblue52138189,fill opacity=0.75,very thin] (axis cs:90.6416666945,0) rectangle (axis cs:92.6388889173333,0);
\draw[draw=whitesmoke238,fill=steelblue52138189,fill opacity=0.75,very thin] (axis cs:92.6388889173333,0) rectangle (axis cs:94.6361111401667,0);
\draw[draw=whitesmoke238,fill=steelblue52138189,fill opacity=0.75,very thin] (axis cs:94.6361111401667,0) rectangle (axis cs:96.633333363,0);
\draw[draw=whitesmoke238,fill=steelblue52138189,fill opacity=0.75,very thin] (axis cs:96.633333363,0) rectangle (axis cs:98.6305555858333,0);
\draw[draw=whitesmoke238,fill=steelblue52138189,fill opacity=0.75,very thin] (axis cs:98.6305555858333,0) rectangle (axis cs:100.627777808667,1);
\draw[draw=whitesmoke238,fill=steelblue52138189,fill opacity=0.75,very thin] (axis cs:100.627777808667,0) rectangle (axis cs:102.6250000315,0);
\draw[draw=whitesmoke238,fill=steelblue52138189,fill opacity=0.75,very thin] (axis cs:102.6250000315,0) rectangle (axis cs:104.622222254333,0);
\draw[draw=whitesmoke238,fill=steelblue52138189,fill opacity=0.75,very thin] (axis cs:104.622222254333,0) rectangle (axis cs:106.619444477167,0);
\draw[draw=whitesmoke238,fill=steelblue52138189,fill opacity=0.75,very thin] (axis cs:106.619444477167,0) rectangle (axis cs:108.6166667,1);
\end{axis}

\end{tikzpicture}
\hfil
\input{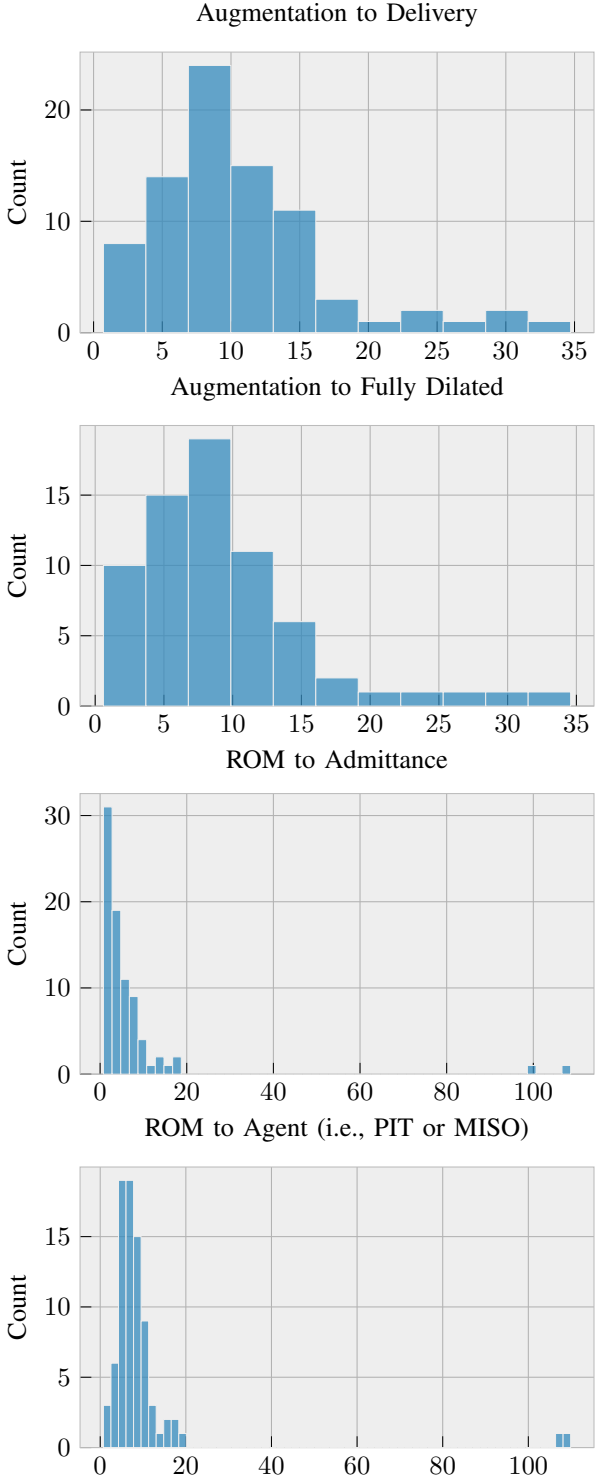}
\caption{The original distribution of the dependent variables for Augmentations and ROMs on the top and bottom rows, respectively. The x-axis in all figures is in hours. }
\label{fig:histograms}
\end{figure}
\begin{figure*}
    \centering
    \includegraphics[width=\textwidth]{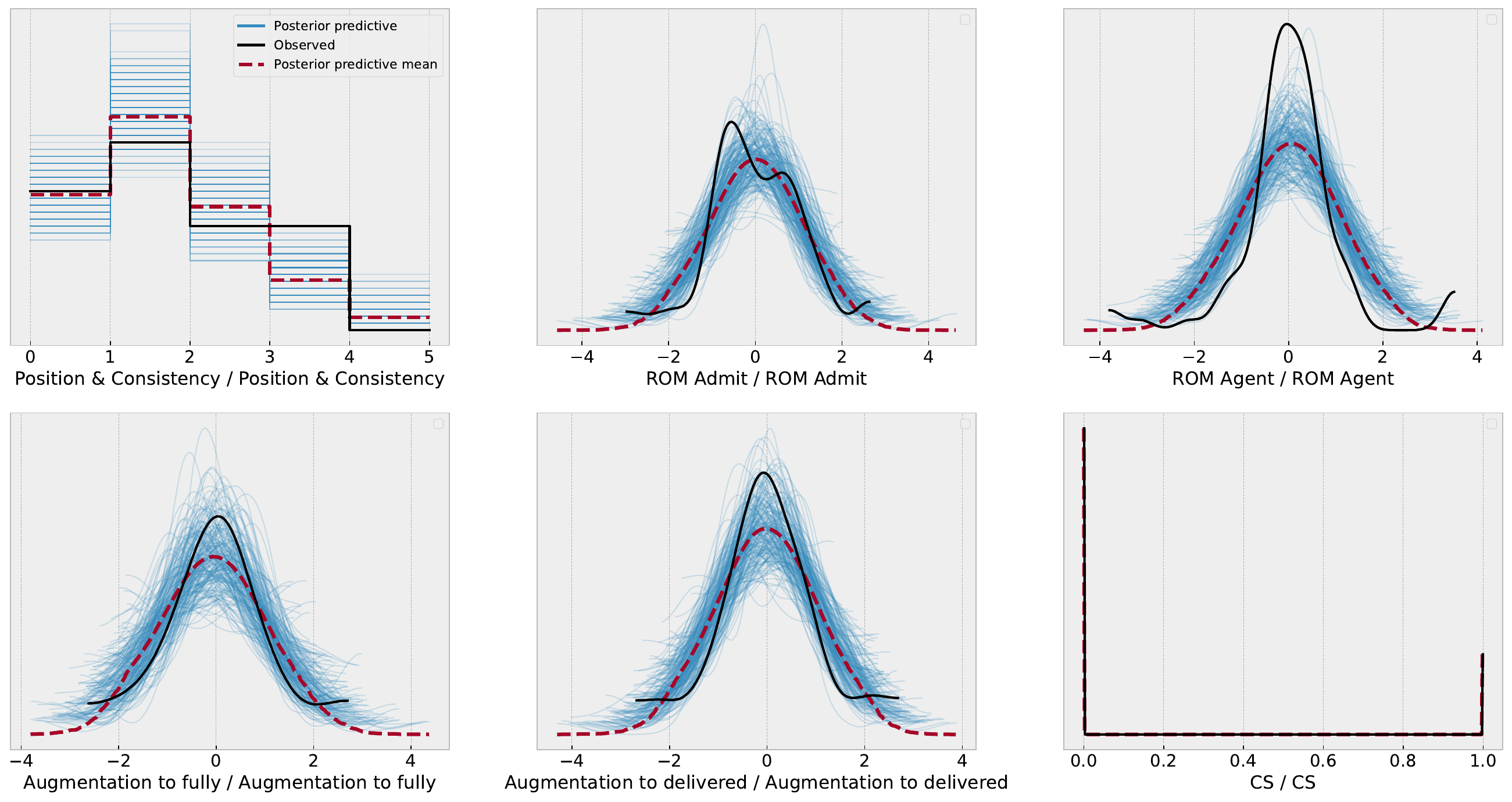}
    \caption{A posterior predictive check is performed to see that the model is appropriate for the observed data. In each case, the dashed red line is the model's mean, the blue lines show one of 200 draws from the distribution to show the variance, and the black line shows the observed data distribution. Our posterior shows high correspondence to the underlying data distribution and thus supports drawing inferences about the population from the results. }
    \label{fig:posteriorPredictiveCheck}
\end{figure*}

A final factor we consider that is not a part of the generative story is a transformation of the ROM and Augmentation outcomes before the parameters are inferred. This is because all variables are positive but with significant skew, as shown in \autoref{fig:histograms}. Both augmentations have a non-trivial tail that extends out to the right, making it hard to explicitly model. This is in part because it is not a heavy-tail in the strict statistical sense, as labor that goes too long will result in a CS, creating a truncated distribution. Similarly, the ROM variables have two patients who are significant outliers, going over 100 hours from rupture to admittance. Though this could be molded with a spike-and-slab prior, the minimal number of samples makes it ineffective and still leaves the asymmetry and skew of the remaining distribution mass a modeling challenge. 

We instead find it much simpler and effective to use a Box-cox transform~\citep{boxCoxTransform} as applied in \autoref{eq:box-cox} to make the outputs appear more Gaussian in shape. 
\begin{equation} \label{eq:box-cox}
    T( y) =\begin{cases}
\left( y^{\lambda } -1\right) /\lambda  & \text{if} \ \lambda \neq 0\\
\ln( y) & \text{if} \ \lambda = 0
\end{cases}
\end{equation}

As part of validating the appropriateness of this model, we perform a posterior predictive check against the six observed features used to compute likelihoods (five outcomes, and the partially imputed PosCon). This is shown in \autoref{fig:posteriorPredictiveCheck}. As can be seen, there is high agreement between our model posterior distribution and that of the true data distribution. The ROM Admit and Agent variables have the lowest fit due to slight bi-modality and a low spread respectively, but are within the general range and shape of the posterior distribution. This provides strong support for the appropriateness of our model to the given data.

\subsection{Implementation Details}

Note, that in extended testing we also tried moving the Box-cox transformation into the Generative model by using the inverse transform before the model prediction. This resulted in identical conclusions as to what we will discuss in the next section and visually indistinguishable posterior predictive checks. For this reason, we prefer to keep it outside the generative story to aid in readability. 

Our model is computed using Markov Chain Monte Carlo~\citep{metropolis1953a} using the NUTS sampler~\citep{hoffman2014a} and implemented with Numpyro~\citep{phan-a} at
with a highest density regions (HDR) of 95\%. For brevity, we will refer to results in/outside the HDR range in terms of ``significance'' due to its greater familiarity with clinical readers whom this interdisciplinary work also targets. We use four chains each with 600 warm-up iterations and then draw 900 samples for inference. We find this is sufficient that the $\hat{R}$ scores for all variables is $\leq 1.005$, indicating convergence \citep{b03a7d33-4a4a-354b-bca4-ded21a5d7dd0}. 

\section{Clinical Results} \label{sec:results}

To start, we first perform a simple Mann-Whitney U test for all given outcome variables of interest. This reflects how current literature would evaluate these factors, ignoring the confounding Bishop scores as noted in \autoref{fig:bishop_process}. In doing so we would reach the same conclusion in our data as the current literature. That Pitocin significantly reduces the time to full dilation and delivery ($p=$ 0.003 and 0.001, respectively). No significant difference would be observed for the ROM times ($p=$ 0.946 and 0.286) or for a CS occurring ($p=$0.300). 
This is not a recommended analysis because it lacks the causal consideration, which we did factor into our Bayesian model.

\begin{figure}[!h]
    \centering
    \includegraphics[width=0.9\columnwidth]{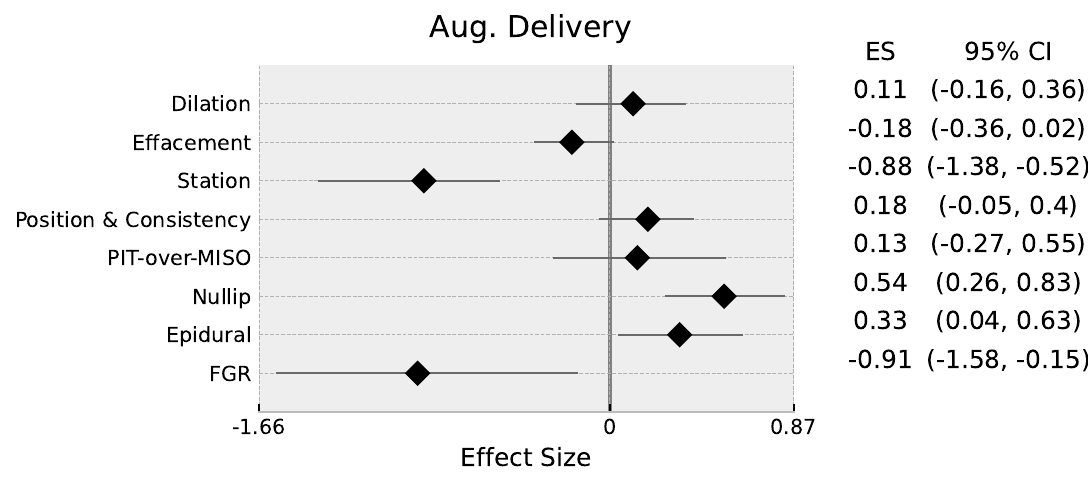}
    \caption{Augmentation to Delivery is the time between when PIT or MISO was started, and when the baby was delivered. The non-difference of PIT-over-MISO is important as it shows the marginal impact of the medication is not as large as previously thought.  }
    \label{fig:aug_delivery}
\end{figure}

\begin{figure}[!h]
    \centering
    \includegraphics[width=0.9\columnwidth]{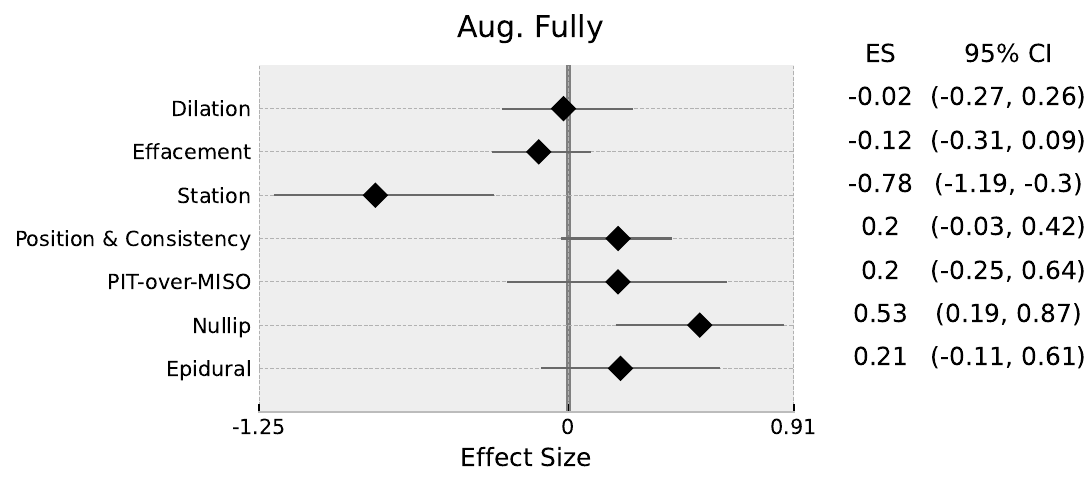}
    \caption{Augmentation to Fully Dilated is the time between when PIT or MISO was started, and when the mother's cervix was fully dilated. Results are consistent with medical understanding, and critically the PIT-over-MISO is non-significant, as it shows the marginal impact is not as large as previously thought.}
    \label{fig:aug_fully}
\end{figure}

According to current literature~\citep{Freret2019}, only PIT should be considered for a larger PROM patient population due to its supposed superiority to MISO in reducing time to delivery. Our results in \autoref{fig:aug_delivery} and \autoref{fig:aug_fully} indicate that this needlessly restricts the provider's options in medication. MISO appears to be equally effective in reducing the time to all outcomes with no increase in adverse events for PROM patients.

In both Augmentation scenarios, a higher station was found to result in a lower time to delivery. This is in line with medical expectations and understanding, as a negative station means the baby has not descended, whereas a positive station is a baby that has descended further into the birth canal. 

Similarly, Fetal Growth restriction (FGR) results in a small baby for delivery. Naturally, a smaller baby is able to pass through the body more easily, and thus the significant reduction in time is in line with expected results. 

In both cases a nulliparous patient, meaning it is their first baby, takes longer to deliver and reach full dilation. This is expected, as the time-to-deliver generally decreases with each subsequent pregnancy. 

Epidurals are known to increase the time required for pushing~\citep{AnimSomuah2018}, and thus potentially delivery time. Simultaneously, Epidurals are known to reduce the time to get to fully dilated~\citep{Wong2005}, where our results show a non-significant result. This is overall consistent with known literature, and in our case, most likely an artifact of limited statistical power given the paucity of patient data.

In its development, a Bishop score was not meant to determine when cervical ripening should be used, it was only meant to determine if induction of labor has an equal chance of successful vaginal birth rate as spontaneous labor. 
Its application to this task is a re-purposing of a useful tool. 
Notably, the Dilation, Effacement, and Position $+$ Consistency components of the Bishop score were not found to be statistically significant in this study. While our power is limited and so non-significance should be taken with appropriate caution, our results may suggest a more refined subset of the Bishop components or alternative scoring could be worth consideration in the future. Indeed, other works have found that the Bishop score is often an ineffective diagnostic for other tasks~\citep{HENDRIX1998,Ivars2016}. 

\begin{figure}[!h]
    \centering
    \includegraphics[width=0.9\columnwidth]{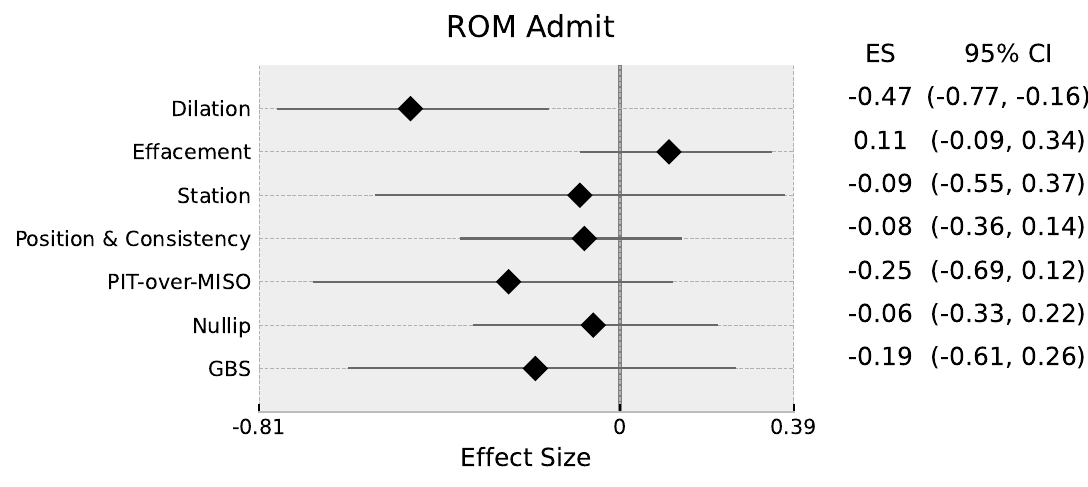}
    \caption{The time between when the Rupture of Membrane occurs and when the mother is admitted to the floor. The results are in line with current medical understanding. }
    \label{fig:rom_admit}
\end{figure}

\begin{figure}[!h]
    \centering
    \includegraphics[width=0.9\columnwidth]{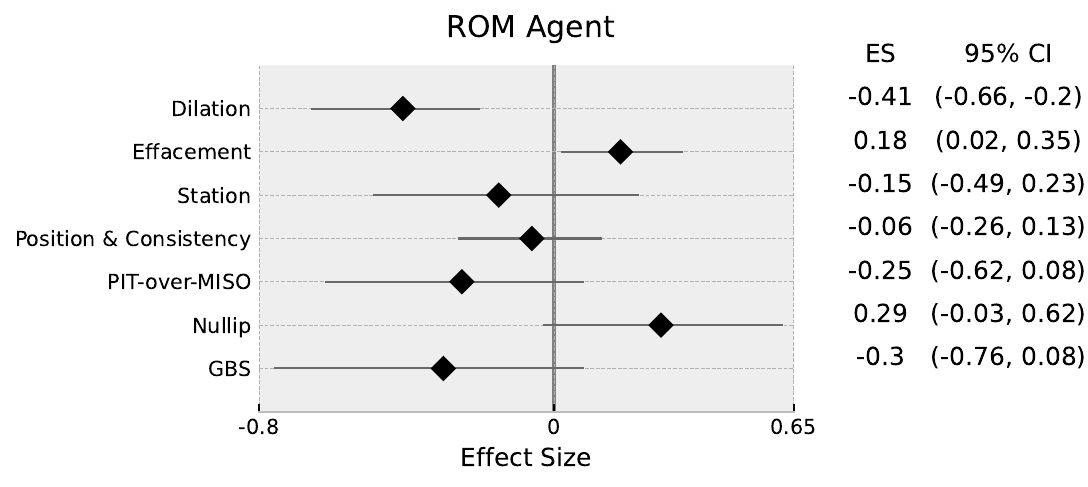}
    \caption{The time between when the Rupture of Membrane occurs and when the PIT or MISO is administered. The results are in line with current medical understanding.}
    \label{fig:rom_agent}
\end{figure}

That an increased dilation results in less time to Admittance in \autoref{fig:rom_admit} is potentially explained by a patient being more uncomfortable, and thus being admitted quicker due to that discomfort. The increased dilation being significant to ROM Agent in \autoref{fig:rom_agent} is more surprising, as our expectation is providers would wait longer to see if delivery progressed without intervention. Though this result has no likely impact on the patient's medical outcome in a clinical context, there may be factors to consider in patient comfort. This would need additional study to determine. 

The other interesting result from  \autoref{fig:rom_agent} is the significance of Effacement, where a more favorable cervix results in a longer time until the agent is administered. Effacement is an indication of a thinner cervix, which is a key requirement for vaginal delivery occurring. For a patient with a thinner cervix, it makes sense for a physician to wait to see if labor begins without intervention. This then results in a longer delay until receiving the Agent. 

\begin{figure}[!h]
    \centering
    \includegraphics[width=0.9\columnwidth]{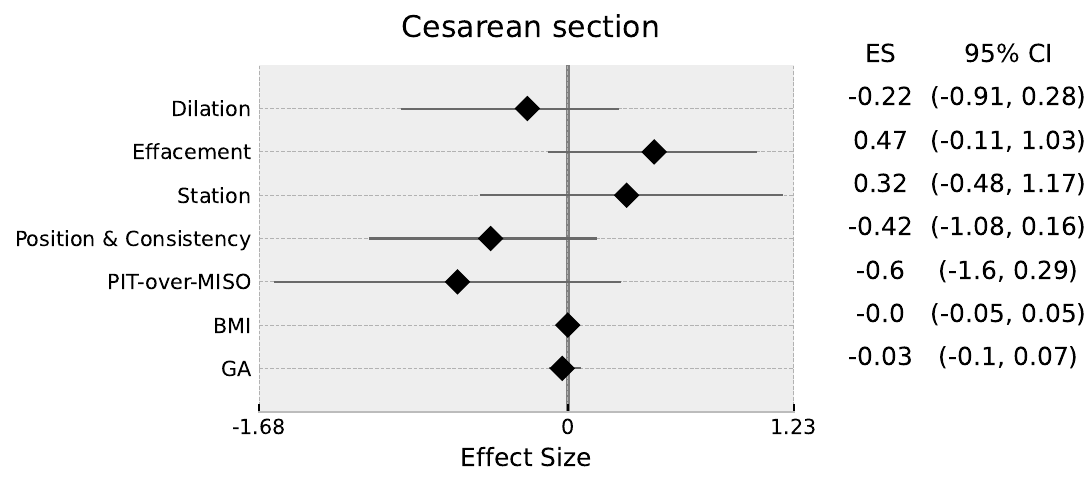}
    \caption{The main adverse event of concern is a Cesarean section because it is a major abdominal surgery. Surgery always carries a higher risk of complications, injury, and possibly death, and so a vaginal birth is preferred. Here we see no difference in the risk of a CS for either drug. }
    \label{fig:cs}
\end{figure}

Finally, \autoref{fig:cs} shows that none of the variables are significant individually for the likelihood of a cesarean section to occur. This is an important positive result because it means, that once confounders are considered, the choice of PIT over MISO does not appear to cause any new risk of needing surgery.

\section{Conclusion} \label{sec:conclusion}

The OBGYN field has been comparatively neglected in adequate research for many decades~\citep{Steinberg2023}. This is being exasperated by a drought of OBGYNs in particular and a decreasing number of physician-scientists to carry out this needed research~\citep{Recanati2022}. Our work shows how a Bayesian approach can be used to make meaningful and clinically relevant inferences from observational data by handling small sample sizes and missing variables when combined with physician expertise. This narrows the scope of needed trials and provides quantified information to practitioners. By demonstrating that Pitocin and buccal misoprostol have similar efficacious with indistinguishable rates of Cesarean section, providers have a larger scope of possible interventions to manage patient care. This informs future randomized control trial designs to focus on including the Bishop factor in the analysis and to further test if misoprostol is still necessary at lower-scores for cervical ripening. 

\section*{Ethics}

This study was carried out under all guidelines and approvals for studying patient chart data at the hospital where it was performed. This work was done in conjunction with professional OBGYNs to evaluate the approach, inform, and confirm the assumptions placed into the model. All patient data will be kept confidential in compliance with local laws.

\bibliography{refs}

\end{document}